%% file: explorviz-vissoft-2026-chatbot.tex
\documentclass[conference]{IEEEtran}
\IEEEoverridecommandlockouts
\usepackage{cite}
\usepackage{amsmath,amssymb,amsfonts}
\usepackage{algorithmic}
\usepackage{graphicx}
\usepackage{textcomp}
\usepackage{xcolor}
\usepackage{balance}
\usepackage[hidelinks]{hyperref}
\usepackage{todonotes}
\usepackage{booktabs}
\usepackage{siunitx}
\usepackage{multirow}
\usepackage{subcaption}
\usepackage{tcolorbox}
\usepackage{tikz}

\newcommand{\figlabel}[1]{%
  \tikz[baseline=(char.base)]{
    \node[
      circle,
      draw,
      inner sep=1pt,
      font=\footnotesize\bfseries
    ] (char) {#1};
  }%
}

\def\BibTeX{{\rm B\kern-.05em{\sc i\kern-.025em b}\kern-.08em
    T\kern-.1667em\lower.7ex\hbox{E}\kern-.125emX}}
\begin{document}

\title{
A Chat Assistant for Software Exploration in a\\ 3D Software Visualization
}

\author{\IEEEauthorblockN{Malte Hansen}
\IEEEauthorblockA{\textit{Department of Computer Science} \\
\textit{Kiel University}\\
Kiel, Germany \\
malte.hansen@email.uni-kiel.de}
\and
\IEEEauthorblockN{Karim Issa}
\IEEEauthorblockA{\textit{Department of Computer Science} \\
\textit{Kiel University}\\
Kiel, Germany \\
stu203382@mail.uni-kiel.de}
\and
\IEEEauthorblockN{Wilhelm Hasselbring}
\IEEEauthorblockA{\textit{Department of Computer Science} \\
\textit{Kiel University}\\
Kiel, Germany \\
hasselbring@email.uni-kiel.de}
}

\maketitle

\begin{abstract}
We present a chat assistant for interactive software exploration, embedded in the 3D software visualization tool ExplorViz.
The assistant builds upon current Large Language Models (LLMs) and enables users to ask questions about the currently visualized software system and trigger actions that change the visualization through natural language.
We integrate the chat assistant in our ExplorViz frontend using the CopilotKit libraries such that probabilistic LLMs are combined with deterministic and tool-based actions similar to implementations using the Model Context Protocol (MCP).
The chat assistant is also enabled to restructure the software system by adding, removing, or modifying parts of the software system in the visualization.

An empirical experiment with eleven participants evaluated both perceived comprehension support and the tool calls that were triggered by the chat assistant.
Participants rated the assistant's generated summaries and explanations as largely correct.
They also reported high usability for actions like highlighting entities in the visualization and the creation of new color themes.

In contrast, open-ended chat-assisted software restructuring in the visualization showed mixed results.
This suggests a need for stronger guardrails for the employed LLM and better incorporation of user feedback.
Overall, the assistant was perceived as usable and promising for reducing interaction overhead during exploratory program comprehension tasks.

We provide a video presenting the chat assistant's use and a reproduction package of the software system that was used in our evaluation.
\end{abstract}


\input{chapters/introduction}

\input{chapters/background}

\input{chapters/related-work}

\input{chapters/implementation}

\input{chapters/visualization}

\input{chapters/evaluation}

\input{chapters/conclusions}

\section{Supplementary Materials}\label{sec:material}
The complete software stack of ExplorViz is available open source on GitHub,\footnote{\url{https://github.com/explorviz}} with complementary documentation available on our website.\footnote{\url{https://explorviz.uni-kiel.de}}
A reproduction package with software artifacts, evaluation results, and a video is available via Zenodo~\cite{hansen2026zenodo}.
In addition, the video is available on YouTube: \url{https://youtu.be/eXSqmo5uDLs}

\section*{Acknowledgments}
We want to thank all participants of our empirical evaluation for their time and valuable feedback.

\clearpage

\providecommand{\doi}[1]{DOI: \href{https://doi.org/#1}{#1}}
\bibliographystyle{myIEEEtran}
\bibliography{bibliography}

\end{document}

%% file: chapters/introduction.tex
\section{Introduction}\label{sec:introduction}
Large language models (LLMs) have become increasingly popular in recent years, particularly for writing-related tasks~\cite{liang2025,liang2025b}.
In addition, various tools using artificial intelligence (AI) and AI-driven approaches using LLMs have been developed for a wide range of disciplines~\cite{yezhu2025,ye2025,liu2025}.
Also, the field of visual analytics has adopted and evaluated the use of LLMs~\cite{xu2024,bendeck2025,brossier2026}.
Recently, the first LLM-driven approaches in the field of software visualization have been presented, for example to generate UML diagrams~\cite{calamo2025,siala2025,vaduva2026}.

Most of the LLM-driven approaches focus on 2D visualizations, and as of now, there is only limited work on the use of LLMs in the context of 3D software visualization tools~\cite{brown2025}.
Traditional computer program interaction and use can be frustrating and time-consuming~\cite{lazar2006}.
Therefore, we consider the use of intuitive, textual, LLM-based interfaces for visualization tools to be a promising way to improve usability.
In this paper, we present the integration of a chat assistant powered by LLMs into our 3D software visualization tool ExplorViz~\cite{fittkau2017, hasselbring2020}.

ExplorViz is a web-based tool that employs the city metaphor~\cite{knight1999,wettel2007}.
The chat assistant is integrated into the frontend of ExplorViz using CopilotKit\footnote{\url{https://www.copilotkit.ai/}} such that the user can interact with it via a chat window that is typical for LLM usage.
Our chat assistant can request data about the state of the visualization as context and modify the visualization by using provided frontend tools.
An empirical evaluation with eleven participants was conducted to generate insights regarding the chat assistant's application for program comprehension, triggering frontend tools, and restructure software systems in the 3D visualization.

The remainder of the paper is structured as follows.
In Section \ref{sec:background}, we introduce our software visualization tool ExplorViz.
In Section \ref{sec:related-work}, we discuss relevant related software visualization and chat assistant approaches.
Section \ref{sec:approach} presents the tool architecture and chat assistant implementation.
Examples of the resulting visualization and interaction with the chat assistant are presented in Section \ref{sec:visualiztion}.
The conducted empirical evaluation is presented in Section \ref{sec:evaluation}.
The paper is concluded with a summary and outlook in Section \ref{sec:conclusion}.

%% file: chapters/background.tex
\section{Background}\label{sec:background}
ExplorViz is our web-based 3D software visualization tool that employs the city metaphor~\cite{fittkau2017,hasselbring2020}.
The use of the city metaphor in software visualization goes back to Knight and Munro, visualizing software as cities in a so-called Software World~\cite{knight1999,knight2000}.
The modern interpretation of the city metaphor is shaped by the tool CodeCity by Wettel et al. who also showed its effectiveness and contributed to the popularity of the metaphor~\cite{wettel2007,wettel2008,morenolumbreras2024}.
CodeCity maps applications, packages, and classes to cities, districts, and buildings in the visualization.

\begin{figure}[htbp]
	\includegraphics[width=\textwidth/2]{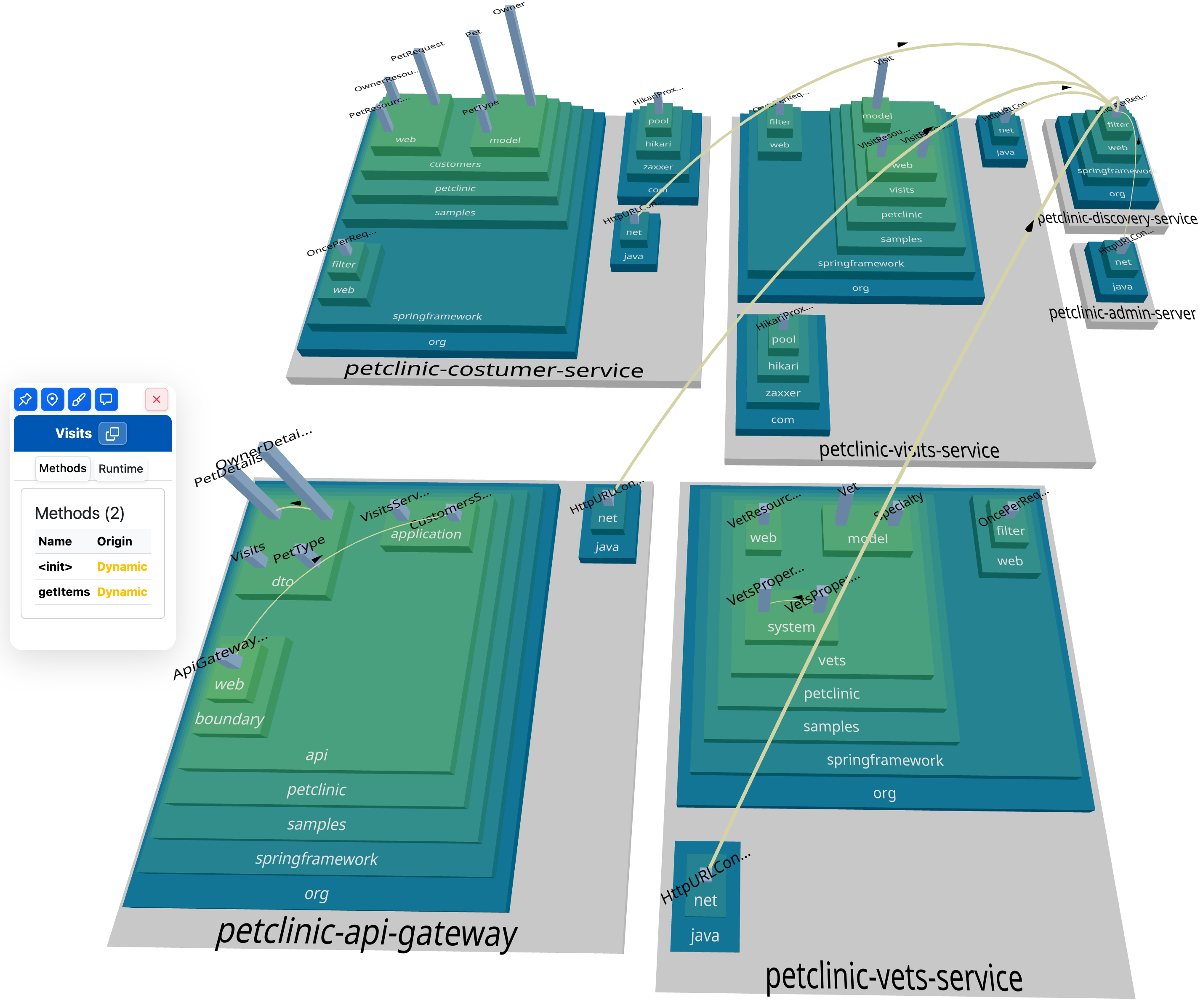}
	\caption{A visualization of the distributed PetClinic in the web interface of ExplorViz.
    Six applications are visualized with a gray foundation and hierarchically stacked packages with a color gradient from blue to green.
    Packages can contain classes. A popup for the class \textit{Vets} is opened. Communication between classes is visualized with yellow arcs.}
	\label{fig:explorviz}
\end{figure}

The visualization of ExplorViz maps applications onto gray city foundations, components (packages) onto hierarchically stacked city districts, and classes onto buildings in the city (see \autoref{fig:explorviz}).
Text labels in the 3D visualization display the names of applications, components, and classes.
The width, depth, and height of buildings can be mapped to metrics such as the number of methods in a class.
The available metrics can also be used to filter the visualization.
Hovering a visual element with the mouse opens a popup with more detailed information, for example displaying methods in a class.
Thereby, we follow the Visual Information-Seeking Mantra by Shneiderman~\cite{shneiderman1996}.

Static program analysis is achieved in a service called Code Analyzer to collect data on a software system's structure, especially regarding its packages and classes.
As the commit history of Git repositories is analyzed in our Code Analyzer, ExplorViz also offers the comparison of different commits, thereby allowing the exploration of code evolution.

In addition to structural data, the focus of ExplorViz lies in the live visualization of execution traces.
Dynamic program analysis is employed to collect runtime data, e.g., with \mbox{NovaTec's} Java agent inspectIT Ocelot\footnote{\url{https://github.com/inspectIT/inspectit-ocelot}} or Kieker~\cite{vanhoorn2012kieker,hasselbring2020kieker,reichelt2025kieker,yang2025kieker}.
The traces are exported to the OpenTelemetry Collector\footnote{\url{https://opentelemetry.io/docs/collector/}} component and need to adhere to the OpenTelemetry\footnote{\url{https://opentelemetry.io}} standard.

The collected traces are aggregated and displayed as yellow arcs between buildings.
These traces represent all method calls between the objects of the associated classes.
Black arrows on the arcs indicate the direction of communication.

The frontend of ExplorViz is written in React, the 3D visualization is rendered onto an HTML canvas in the browser using three.js.\footnote{\url{https://threejs.org/}}
To bridge the gap between React and three.js, we use the React Three Fiber\footnote{\url{https://docs.pmnd.rs/react-three-fiber}} library such that every 3D object in the scene can be wrapped as a declarative React component.
The state of the visualization is managed using the Zustand\footnote{\url{https://zustand-demo.pmnd.rs/}} library.
Visualization settings are persisted in the browser storage.

%% file: chapters/related-work.tex
\section{Related Work}\label{sec:related-work}

\paragraph{Chat assistants in software engineering} The emergence of conversational LLM tools (e.g., ChatGPT, Claude) has spurred research on chat assistants for code comprehension.
A study by Lehtinen et al.~\cite{lehtinen2024} examined ChatGPT answering natural-language questions about programs.
Participants rated the correctness of explanations highly (mean $\sim$4.5/5), though some hallucinations were noted.
Richards and Wessel~\cite{richards2024} propose adding a “Theory of Mind” layer to personalize answers to the user’s expertise, showing the potential to tailor assistance.
AlOmar et al.~\cite{alomar2025} analyze ChatGPT in code refactoring dialogues, finding that developers rely on it for maintainability suggestions, but sometimes need to write prompts to get concise results.
In summary, chat-based AI can generate useful summaries and refactorings (often using affirming language), but consistency and precision are concerns~\cite{lehtinen2024, alomar2025}.

\paragraph{Tool-augmented LLM agents} To combine LLM reasoning with concrete actions, several frameworks have been developed.
Platforms like Anthropic’s Model Context Protocol (MCP)~\cite{anthropic2024mcp,anthropic2025mcp} standardize how LLMs query data sources and invoke tools via a common API.
The CopilotKit libraries (used in our system) similarly allow embedding LLM agents into frontends, akin to MCP, to issue deterministic UI actions.
Most closely related to our setting, Brown and Mulder extend an immersive virtual-reality environment for binary reverse engineering with an integrated LLM agent that can query binary analysis tools, answer technical questions, and dynamically generate immersive 3D visualizations aligned with analyst tasks~\cite{brown2025}.
Their pilot study reports that while the LLM can generate meaningful 3D call graphs for small programs that align with design principles, the quality of the generated output varies widely, leading them to raise open questions about whether LLMs can reliably function as visualization agents that construct 3D representations, reflecting cognitive design principles without explicit training~\cite{brown2025}.

In a related agentic setting, V\u{a}duva et al.~\cite{vaduva2026} present Code2UML, in which a hierarchy of five specialized LLM agents, combined with a deterministic context-engineering layer that compacts intermediate code representations to fit model context limits, generates UML diagrams from large code repositories.
Across 12 repositories and several diagram types, they report high syntactic validity, while deliberately prioritizing architecturally important entities over exhaustive coverage.
This illustrates how multi-agent decomposition and context engineering can address the scalability limits of single-prompt LLM visualization.

These findings directly inform our design.
Like Brown and Mulder, we treat the LLM as an actor \emph{on} the visualization rather than a passive chat window, mapping natural-language requests to concrete view operations in ExplorViz.
Unlike their binary-analysis setting, where the agent synthesizes new 3D representations, our software-city target lets us bind agent actions to well-defined existing entities (classes, packages, communication links), which constrains the agent's output and makes the resulting actions easier to confirm and undo.
Our integration of CopilotKit with ExplorViz thus follows this trend of tool-augmented visualization agents, but our findings echo their observation that output quality is uneven: guardrails and explicit user confirmation remain important to keep agent-driven changes transparent and to prevent unintended modifications~\cite{alomar2025}.

\paragraph{Program comprehension and usability evaluations} General research on program comprehension has long emphasized tools that reduce cognitive load~\cite{dealwis2007,shargabi2020,husak2025}.
3D visualizations leverage spatial reasoning and overviews of software architecture~\cite{wettel2008, hasselbring2020}.
Our chat assistant aims to further lower barriers by answering user queries naturally.
Evaluations of comprehension tools typically measure correctness and usability.
For instance, ExplorViz’s baseline studies reported positive effectiveness in collaborative scenarios~\cite{krauseglau2022}.
In AI-assisted settings, researchers measure correctness of answers (often via Likert ratings) and workflow efficiency.
Lehtinen et al.~\cite{lehtinen2024} found high user satisfaction with GPT-4 explanations.
While LLMs can augment comprehension, interactive agent-driven changes require stronger feedback and verification steps.

%% file: chapters/implementation.tex
\section{Chat Assistant Implementation}\label{sec:approach}
In this section, we present the tool architecture, initial instructions that our chat assistant receives, the tools that our chat assistant implementation can trigger in ExplorViz, the application state made available as context, and an illustration of the user-facing chat interface.

\begin{figure}[htbp]
	\includegraphics[width=\textwidth/2]{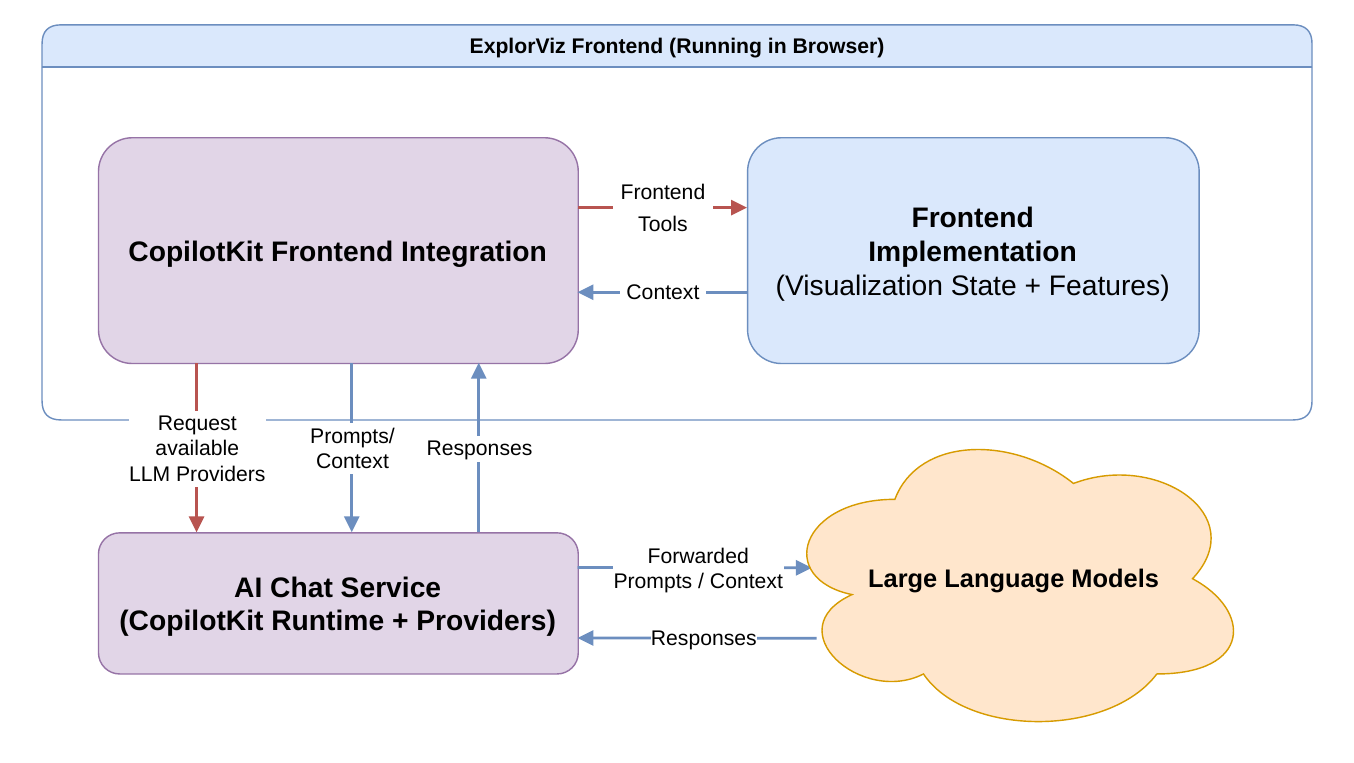}
	\caption{Overview of chat assistant architecture using \mbox{CopilotKit}. Red arrows indicate relevant control flow while blue arrows illustrate data flow.}
	\label{fig:architecture}
\end{figure}

\textbf{Architecture}
Our goal is to provide users with a chat assistant that can access and manipulate the visualization state.
As much of the visualization state is managed directly in the frontend, the implementation of a separate MCP server is not a good fit for our overall architecture.
We decided to employ parts of the CopilotKit software stack to achieve a direct integration into our React frontend.
The resulting architecture for our chat assistant is presented in \autoref{fig:architecture}.

We use the CopilotKit React dependencies in the frontend to set initial instructions, register frontend tools, and register available context interfaces for our chat assistant.
Direct communication from within the frontend to an LLM provider is not feasible.
Therefore, we implemented an AI Chat Service that manages required API keys via environment variables and acts as a bridge between our frontend and the currently selected LLM.
The API keys for LLMs need to be provided to the AI Chat Service when it is started.
The AI Chat Service is a Node.js server that includes dependencies for the CopilotKit Runtime\footnote{\url{https://docs.showcase.copilotkit.ai/backend/copilot-runtime}} and dependencies to provide access to different LLM providers.
The following LLMs from OpenAI, Google, and Anthropic are registered with their model ID in the AI Chat Service:
\begin{itemize}
    \item OpenAI: gpt-5, gpt-5-mini, gpt-4.1, gpt-4.1-mini, gpt-4o, gpt-4o-mini
    \item Google: gemini-3-pro-preview, gemini-2.5-pro, gemini-2.5-flash, gemini-2.5-flash-lite
    \item Anthropic: claude-sonnet-4-5, claude-haiku-4-5, claude-opus-4-5, claude-opus-4-1
\end{itemize}
The frontend is provided with the list of models for which an API key is provided in the backend.
After this, prompts and context as well as responses containing instructions for tool calls are forwarded by the AI Chat Service between the frontend and LLM.
This separation keeps LLM provider credentials outside the browser while allowing the visualization to execute tool calls locally without calls to a backend service.

\textbf{Instructions}
As we use general-purpose LLMs, it is necessary to add some contextual information about ExplorViz and the chat assistant's purpose at the beginning of a chat session.
This is done in the background as so-called instructions via CopilotKit.
Iterative testing during the development led us to the following instructions for the chat assistant:

\textit{
``You are assisting the user as best as you can. You are embedded inside the frontend of a software called ExplorViz. ExplorViz is a tool for visualizing and analyzing software landscapes in 3D. The user can interact with the 3D visualization by asking you to highlight, open, or close components. You can also provide information about the applications, packages, classes, and methods present in the landscape. Applications are the highest level of organization in the landscape, containing packages. Packages can contain sub-packages and classes. Classes can contain methods and interfaces. By default all components are open (visible) and not highlighted.''
}

\textbf{Context for the Chat Assistant}
In order to make information about the visualization's current state in ExplorViz accessible to the chat assistant, CopilotKit enables the registration of application state as so-called agent context.
Via a React hook, the returned data of functions in ExplorViz is given as context to the LLM in the background.

When designing the interfaces for context retrieval in \mbox{ExplorViz}, we followed three guiding principles:
\begin{enumerate}
    \item Relevance for available frontend tools: The context should be aligned with the available tool calls for the chat assistant such that the chat assistant can make informed decisions about which calls should be made to fulfill a user request.
    \item Small data size: The provided data needs to be compact enough to fit into the LLM's context window and should be as small as possible in order to reduce token usage and thus cost of using the chat assistant.
    \item Stable across renders: The requested data should not change automatically without a user action, thereby helping the chat assistant to give up-to-date and correct responses.
\end{enumerate}

Principles two and three are the reasons that we left out communication (traces) from the context: Especially for large software systems, communication data can become quite large and, when live tracing is enabled in ExplorViz, this data updates frequently.

The following interfaces are implemented via React Hooks to provide context for our chat assistant:

\textit{Application Summary}
List with summary about all applications, contains application ID, application name, programming language, origin of data (static/dynamic program analysis), package count, class count, and method count.

\textit{IDs of Closed Components}
List of identifiers (IDs) that indicate which components are closed.
Important for the chat assistant to know which components and classes are visible and thus, which tool calls for opening or closing components are sensible.

\textit{IDs of Highlighted Entities}
List with highlighted IDs such that a user may ask questions about highlighted entities and such that the chat assistant can consider this list when triggering highlights itself.

\textit{ID of Hovered Entity}
ID of the entity that is currently hovered by the user's cursor, else returns null. Enables a user to ask questions about the currently hovered entity.

\textit{IDs of Removed Entities}
List with IDs of already removed entities such that the chat assistant knows which elements are no longer part of the visualization due to restructuring actions.

\textit{Can Undo / Can Redo}
Boolean values that indicate if, according to the restructuring changelog, restructuring actions can currently be undone or redone.

\textit{Visualization Settings} 
Flattened map with current settings for the visualization.
These include the values for colors and layout options as well as the ability to toggle and configure additional visual features such as semantic zoom or the mini map.
For all settings, an ID, value, and description are exposed.

\textit{Default Visualization Settings}
Flattened map with default settings such that the chat assistant has a point of reference to adapt the values of settings.

\textit{Sidebar States}
Two context endpoints that indicate with a boolean value if the left or right sidebar is opened.

\textit{Opened Sidebar Components}
Two context endpoints that indicate which React components are opened in the sidebars.
At most one component can be opened per sidebar.
The left sidebar contains components for search, filtering, and animated trace replay.
The right sidebar contains components for collaboration controls, the chat window, Visual Studio Code extension, restructure interface, visual snapshots, and settings.

\textbf{Frontend Tools}
To enable the chat assistant to trigger changes in the visualization, it has access to several tools that can be called via frontend functions.
These tool calls also include software restructuring features based on the work of Krause-Glau et al.~\cite{krauseglau2024}.
The restructuring calls only alter the frontend state regarding the software landscape and result in an updated visualization.
Since the source code is generally unavailable for software systems observed via OpenTelemetry, code changes in the observed software system are not implemented.
However, the generated changes are tracked in a changelog and can be exported to a GitLab issue to preserve the restructuring recommendations.

The implemented frontend tools are listed below with short descriptions.

\textit{Query Landscape Data} Allows querying data on some or all applications with one of the following detail levels: application summary, package-level, class-level, or method-level.
Also provides a parameter to limit the length of the results.
This action complements the generic Application Summary context with the ability to query specific and filtered data about the software landscape.

\textit{Highlight}
Changes the color of an entity to the highlighting color that is configured in the settings.

\textit{Ping Entity}
Temporarily places an animated visual indicator above an entity. Can be used as an alternative to the highlighting of entities.

\textit{Open/Close Component}
Given a component ID and a boolean value, opens or closes a component.

\textit{Focus Camera}
Given an entity ID, the camera is moved such that the specified entity is centered on the screen (e.g. centered on a building representing a class).

\textit{Reset Camera Position}
Resets the camera position such that the visualized software system is centered on the canvas.

\textit{Open/Close Sidebar}
Two actions that allow opening or closing the left or right sidebar, respectively.

\textit{Open/Close Sidebar Component}
Two actions that allow opening or closing a React component, indicated by a boolean value, in the left or right sidebar, respectively.

\textit{Filter Entities}
Opens the React component for entity filtering in the left sidebar and applies filter options, such as a minimum threshold for method count.

\textit{Update Setting}
Given a setting ID and a new value, updates a setting for the visualization. Can also be used to reset a single or all settings.

\textit{Add Application}
Given a name, adds a new application to restructure the given software system.
Optionally allows a list of classes with fully qualified names to add components and classes, too.

\textit{Add Classes to Application}
Takes a list of classes containing fully qualified names and method names to add components and classes to an existing application.

\textit{Remove Entity}
Given an ID, removes an application, component, or class as part of a software restructuring.

\textit{Undo/Redo Edit}
Two actions that enable the chat assistant to undo or redo the previously listed restructuring actions.

\textit{Take Screenshot}
Captures a screenshot of the whole page that is scaled down to 1280x720 pixels for improved compatibility with LLMs.

\textit{Click on Screen}
Enables the chat assistant to trigger a single click event anywhere in the ExplorViz interface. Coordinates are given as normalized parameters between zero and one for both axes.
The description of this tool instructs the LLM to take a screenshot as a reference before clicking on the screen.

\begin{figure}
\centering
\begin{tikzpicture}
  \node[anchor=south west,inner sep=0] (img)
    {\includegraphics[width=\linewidth]{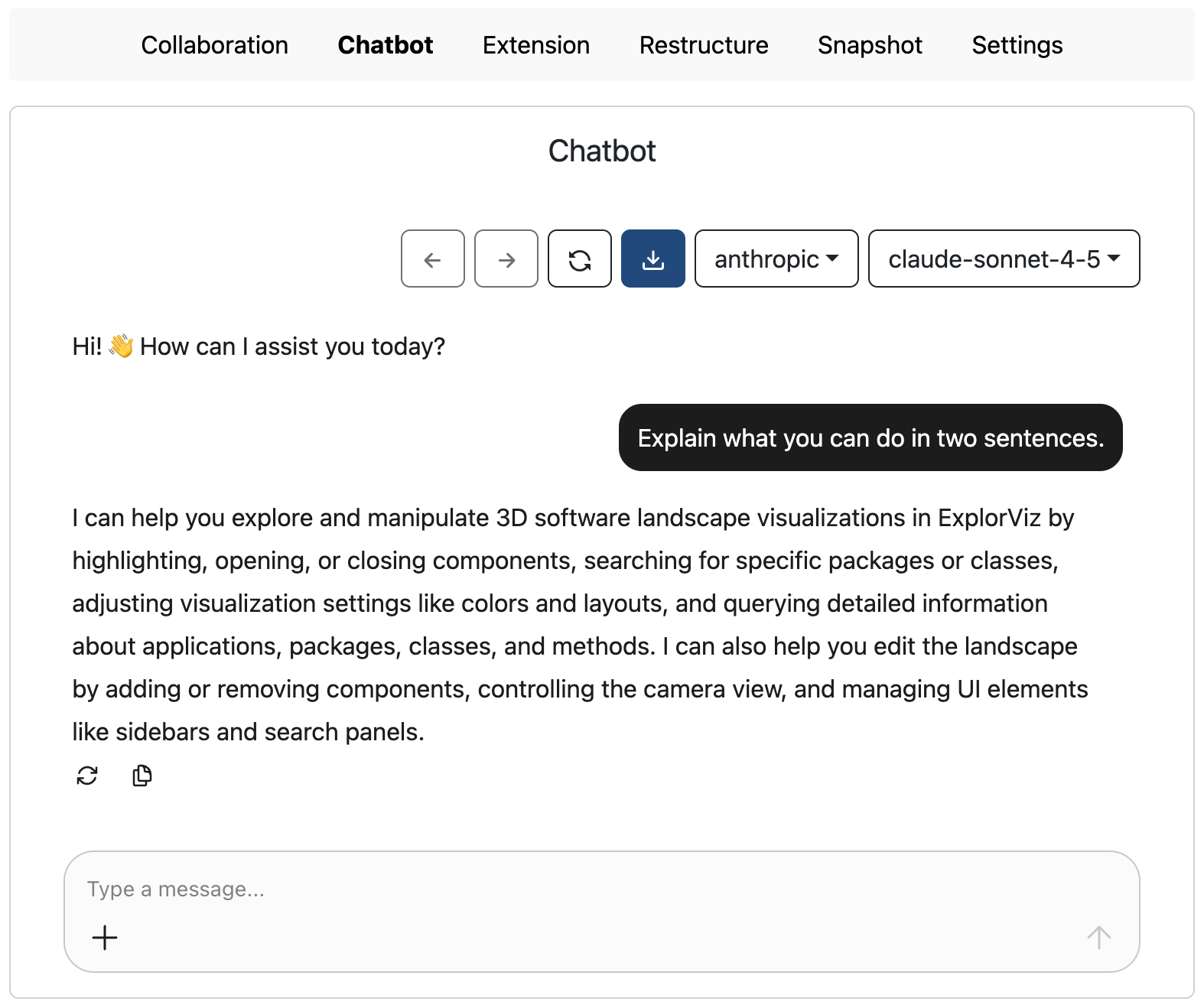}};

  \begin{scope}[x={(img.south east)},y={(img.north west)}]
    \node at (0.325,0.995) {\figlabel{A}};
    \node at (0.39,0.805) {\figlabel{B}};
    \node at (0.48,0.805) {\figlabel{C}};
    \node at (0.54,0.805) {\figlabel{D}};
    \node at (0.64,0.805) {\figlabel{E}};
    \node at (0.82,0.805) {\figlabel{F}};
    \node at (0.0959,0.195) {\figlabel{G}};
    \node at (0.135,0.075) {\figlabel{H}};
  \end{scope}
\end{tikzpicture}
\caption{The chat assistant (called Chatbot in the UI) exposes a chat interface in the right-hand sidebar of the ExplorViz frontend.}
\label{fig:chat}
\end{figure}

\textbf{Chat Interface} 
The chat interface is exposed via a React component that can be opened in the sidebar on the right side of the interface of ExplorViz (see \autoref{fig:chat} \figlabel{A}).
Via arrow buttons (see \autoref{fig:chat} \figlabel{B}), the user can step backwards and forwards in the chat history.
By default, all previous chat messages are sent alongside a user prompt as additional context to achieve conversational awareness.
The user can clear the chat, removing old chat messages, by the click of a button (see \autoref{fig:chat} \figlabel{C}).
Furthermore, a JSON transcript of the chat can be downloaded (see \autoref{fig:chat} \figlabel{D}).
This transcript does not only contain the text messages that are visible to the user, but also additional metadata and tool calls including the tool call's results.
Based on available API keys in our implemented backend service, the user can select an LLM provider (see \autoref{fig:chat} \figlabel{E}) and a specific LLM (see \autoref{fig:chat} \figlabel{F}).
Responses of the chat assistant can be regenerated or copied via buttons at the end of a response (see \autoref{fig:chat} \figlabel{G}).
In addition to textual input, the upload of files such as images is supported (see \autoref{fig:chat} \figlabel{H}).

%% file: chapters/visualization.tex
\section{Visualization Example}\label{sec:visualiztion}
\begin{figure*}[t]
\centering
\begin{tikzpicture}
  \node[anchor=south west,inner sep=0] (img)
    {\includegraphics[width=\textwidth]{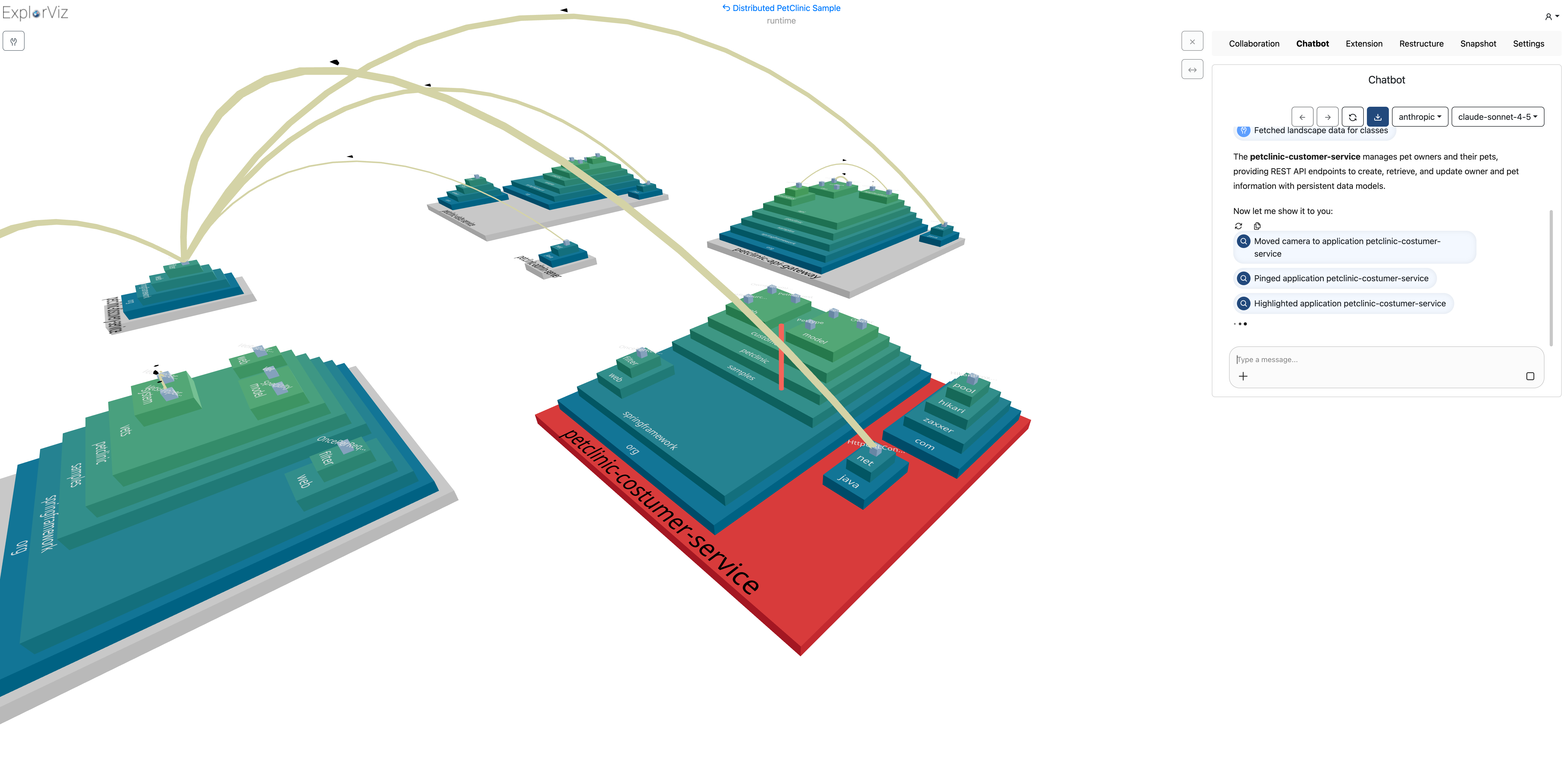}};

  \begin{scope}[x={(img.south east)},y={(img.north west)}]
    \node at (0.77,0.82) {\figlabel{A}};
    \node at (0.77,0.68) {\figlabel{B}};
    \node at (0.77,0.58) {\figlabel{C}};
    \node at (0.51,0.10) {\figlabel{D}};
    \node at (0.50,0.46) {\figlabel{E}};
  \end{scope}
\end{tikzpicture}
\caption{Visualization of the distributed PetClinic with the chat assistant open. The chat assistant was asked to explain the purpose of customer service, after which it should show the service by highlighting and pinging it, as well as moving the camera to it.}
\label{fig:visualization}
\end{figure*}

An exemplary visualization of the distributed version of the PetClinic\footnote{\url{https://github.com/spring-petclinic/spring-petclinic-microservices}} with our chat assistant is presented in \autoref{fig:visualization}.
For this, we started a new chat with the following prompt: \textit{``Explain what the customer service does in one sentence, then show me the service (using camera, highlighting, and pinging).''}

Transparently presented to the user, it can be seen that the chat assistant first queries the available landscape data (see \autoref{fig:visualization} \figlabel{A}).
Based on the queried data, the response with a description of the requested service is generated.
Subsequently, the chat assistant makes calls to the frontend tools to move the camera, ping the service, and highlight the service (see \autoref{fig:visualization} \figlabel{B}).
Animated dots indicate that the chat assistant is not yet done generating a response (see \autoref{fig:visualization} \figlabel{C}), as the chat assistant usually generates a brief summary at the end of a response.
In the resulting visualization, the virtual camera is focused on the customer service.
In order to highlight the service, the city foundation is colored red (see \autoref{fig:visualization} \figlabel{D}) and the center of the visualized application is temporarily pinged (see \autoref{fig:visualization} \figlabel{E}).

%% file: chapters/evaluation.tex
\section{Evaluation}\label{sec:evaluation}
In order to gain insights into the practical application of our implemented approach, we conducted an empirical evaluation in December 2025.

\subsection{Research Questions}
This evaluation is driven by three main areas of interest:
\begin{enumerate}
    \item \textbf{Program Comprehension} 
What are the perceived benefits and limitations that arise from a chat assistant that has access to software visualization data in terms of program comprehension?
    \item \textbf{Tool Execution} What are the advantages and disadvantages of giving a chat assistant, triggered by natural language, the ability to execute tool calls in an otherwise deterministic software visualization? How does it compare to manual interaction with the user interface in terms of perceived effort, usability, and perceived correctness?
    \item \textbf{Chat-Based Software Restructuring} What are the advantages and disadvantages of AI-assisted software restructuring actions that are executed as part of a software visualization? How well does the outcome align with the user's intent and what improvements could be made?
\end{enumerate}

\subsection{Setup}
The study was conducted in-person in the computer laboratory of the Software Engineering Group at Kiel University.
A desktop computer with two monitors was used: one main monitor and one monitor for displaying the survey.
The computer was running Debian GNU/Linux 13 and had the following hardware specifications: Intel Core i5-6500 CPU, 16GB DDR4 RAM, and an Nvidia GTX 1070 GPU.
The main monitor was 34 inches with a resolution of 2560 x 1080 pixels. The monitor used for the survey was placed to the left of the main monitor and was 24 inches with a resolution of 1920 x 1200 pixels.

The PetClinic Sample application's distributed version\footnote{\url{https://github.com/spring-petclinic/spring-petclinic-microservices}} was used as the software system for evaluation.
The following Docker images with the prepared software landscape in the mocked backend (called Demo Supplier) were running on the computer: \textit{explorviz/demo-supplier:chatbot-study-2025}\footnote{\url{https://hub.docker.com/r/explorviz/demo-supplier/tags}}, \textit{explorviz/ai-chat-service:chatbot-study-2025}\footnote{\url{https://hub.docker.com/r/explorviz/ai-chat-service/tags}}, and \textit{explorviz/frontend:chatbot-study-2025}\footnote{\url{https://hub.docker.com/r/explorviz/explorviz-frontend/tags}}.
The frontend of ExplorViz was accessed with Firefox\footnote{\url{https://www.firefox.com/}} version \textit{140.6.0esr}.

\subsection{Participants}
Eleven people, who were mostly computer science students, took part in the evaluation.
At the start of the evaluation, participants of the evaluation were asked to rate their proficiency in programming, how well they know ExplorViz, and how frequently they use LLM tools.
The results are presented in \autoref{tab:participants}.
Participants were also asked which LLMs and AI tools they had used and for which tasks.
ChatGPT stood out as it had been used by all participants.
The most prominent use cases were code generation/refactoring and debugging, reported by nine and eight participants, respectively.

\begin{table}[h]
\centering
\begin{tabular}{|c|c|c|c|}
\hline
\textbf{Question}         & \textbf{Mean} & \textbf{SD} & \textbf{Range} \\ \hline
Programming Experience    & 4.00 & 0.77 & 3-5 \\ \hline
Experience With ExplorViz & 2.82 & 0.98 & 1-4 \\ \hline
Frequency of LLM Usage    & 3.91 & 1.14 & 2-5 \\ \hline
\end{tabular}
\caption{Results on self-reported experience of participants with computed mean values, standard deviation, and range of given answers. Answers were given on a Likert scale from 1-5 (higher numbers mean more experience / more frequent usage).}
\label{tab:participants}
\end{table}

\subsection{Methodology}
The workspace was prepared for participants such that ExplorViz was running on the main monitor and a questionnaire with instructions in LimeSurvey (licensed to Kiel University) was opened on the second monitor.
The chat assistant was configured to use the \textit{Claude Sonnet 4.5} model.
The models from Anthropic have shown the best compatibility with our implemented features such as processing screenshots, interweaving tool calls and explanations seamlessly, and a good response time.
Among the three available Claude models (Haiku, Sonnet, and Opus), we opted for Sonnet to achieve a good balance of quality, response time, and token usage (cost).
We used the standard server tier of Claude\footnote{\url{https://platform.claude.com/docs/en/api/service-tiers}} and had the standard license for individual users (no academic license / no discount).

In the beginning, participants were informed about the purpose of the study and that their participation is voluntary.
Additionally, participants were informed that they had the chance to win one of three €15 Amazon gift cards that were randomly distributed to three participants after the evaluation was concluded.

Afterwards, the participants provided some data on their background (see \autoref{tab:participants}).
The main part of the evaluation consisted of the following tasks:

\begin{enumerate}
    \item \textit{Tutorial}
Participants completed a brief tutorial covering navigation in the 3D space, entity inspection, highlighting, searching, filtering, and adjusting settings to ensure familiarity with the core features of ExplorViz.
No data was collected for this introductory task.
    \item \textit{Application-Level Program Comprehension}
Participants were tasked to inspect the \texttt{petclinic-vets-service} of the distributed PetClinic by manually using features of ExplorViz, e.g. by inspecting the package names and data displayed in popups.
Based on this, they should give a brief description of the service.
Afterwards, they were asked to prompt the chat assistant for a service summary and rated how much additional understanding they gained and how correct the response appeared to them.
    \item \textit{Program Comprehension}
Participants were asked to describe what a package with the name ``dto'', based on the contained classes, likely is responsible for in the distributed PetClinic.
The participants should also rate how confident they are with their answer.
The package in question contains classes with data transfer objects (DTOs), a concept that is not covered in mandatory undergraduate computer science courses at Kiel University.
The participants were then instructed to ask the chat assistant for an explanation and rated the information gain, perceived correctness, and whether the concept “DTO” was explained.
    \item \textit{Tool Use}
In this task, we want to compare the manual use of tools with tool calls triggered by the chat assistant.
Participants were tasked to manually highlight all model classes in the visualization, i.e. classes that contain the string ``model''.
The participants were encouraged to use the search function of ExplorViz.
Then they were asked to make use of the chat assistant to perform the same operation and rated perceived correctness, ease of use, and classified possible mistakes of the chat assistant.
    \item \textit{Configuration of Settings}
Participants should manually create a color theme in the settings and then ask the chat assistant to generate different color theme variants.
The participants rated how well it matched their expectation and the relative ease of use.
    \item \textit{Software Restructuring}
Participants should prompt the chat assistant to improve or change the structure of the displayed software landscape.
They rated explanation quality, overall perceived correctness, how well it matched their expectations, and whether additional confirmation prompting was required.
No manual software restructuring was done in this task, as it would have been too time consuming for our evaluation.
\end{enumerate}

During the tasks, the participants were usually asked to rate their experience on a five-point Likert scale and had the option to add additional remarks in free text fields.
At the end of the survey, a post questionnaire asked for general feedback on our implemented approach.
This included the rating of response times, overall ease of use, and overall experience of interacting with the chat assistant.

\subsection{Results}
\begin{figure*}
  \includegraphics[width=\textwidth]{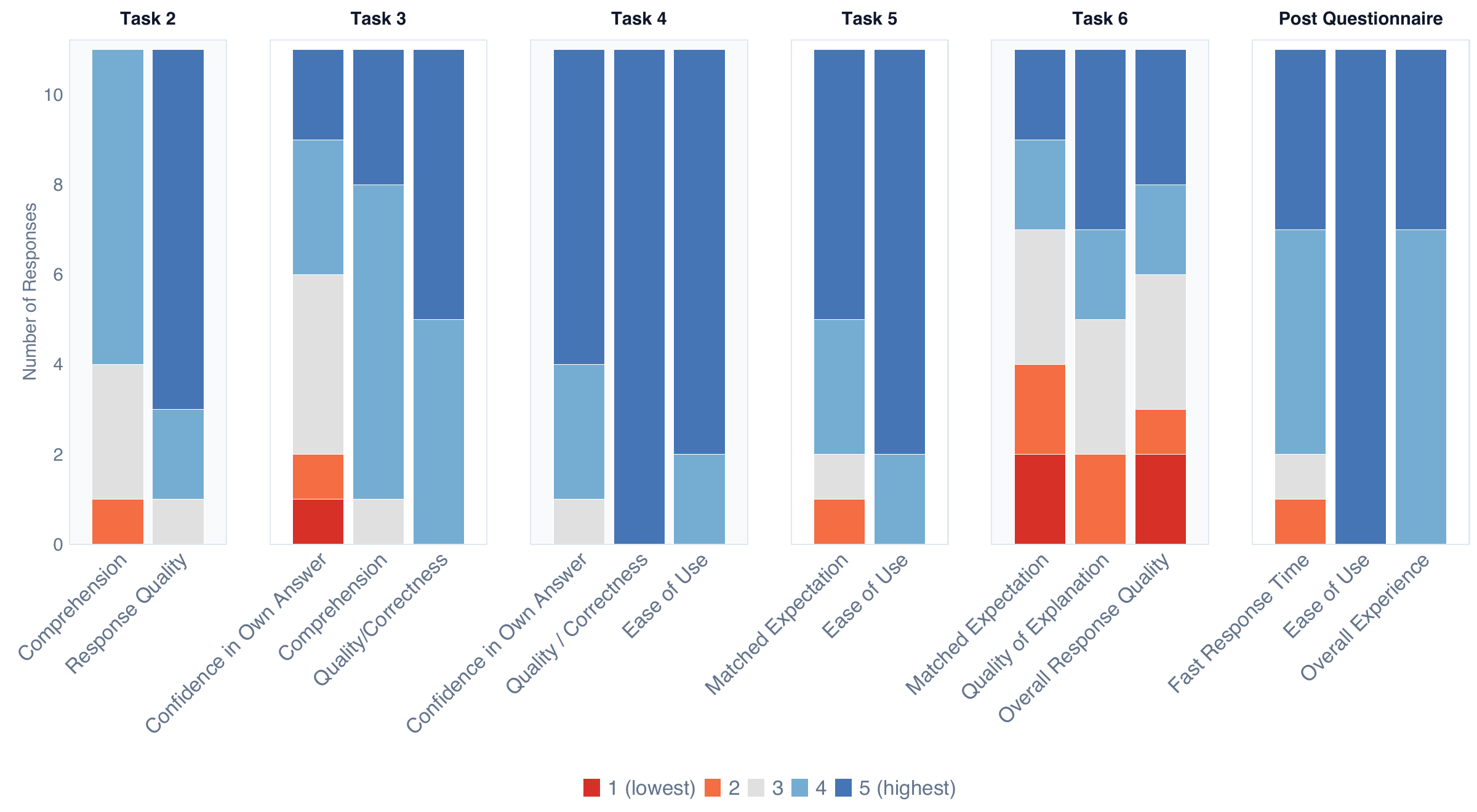}
  \caption{Overview of quantitative results on the five-point Likert scale questions for Tasks 2 through 6 and the post questionnaire.}
  \label{fig:evaluation}
\end{figure*}

In this subsection, we present the results of our evaluation, grouped by tasks.
Task 1 was the introduction to ExplorViz and thus will be omitted here.
The quantitative results of questions using a five-point Likert scale are presented in \autoref{fig:evaluation}.
The participants also had the option to give several textual comments.
We can only mention a very limited selection of comments here, but all results of our evaluation are available in the provided reproduction package (see Section \ref{sec:material}).

In Task 2, all but one participant, who did not type in an answer, recognized that the service manages data about veterinarians.
The additional descriptions resulting from the manual inspection varied in detail, but all participants reported having gained more information from the chat assistant's answer.
The comments also address the level of detail in the responses, as one participant reports that the chat assistant provided ``much more detailed and user friendly'' information.
Another participant noted: ``Some more in-depth information, but nearly too technical (much) for a brief summary''.
Regarding the response quality, eight people rated it as very good and two people as good (see \autoref{fig:evaluation}).
With a numerical mapping from one to five for the Likert scale, the response quality averaged at 4.64/5 points.

In Task 3, the participants are asked to briefly explain what a package called dto likely does.
All participants recognize that it contains classes to store data or models.
However, only three out of eleven participants explicitly use the term ``data transfer object'' in their answers.
The confidence in their own answers is mixed (see \autoref{fig:evaluation}).
After the participants prompted the chat assistant for additional information, all participants indicated having received additional information, averaging at 4.18/5 points.
All participants also stated that the chat assistant explained the term DTO to them.
The perceived correctness/quality of the chat assistant's answer was rated with 4.55/5 points.
However, in the textual comments, two participants mentioned that correctness is difficult to quantify, for example: ``Without deep knowledge of the software the correctness is hard to asses but the superficial information provided seems to be correct''.

In Task 4, the participants were asked to highlight all model classes.
Only one participant reports a medium confidence for this manual task, while all other participants report high or very high confidence (see \autoref{fig:evaluation}).
As the focus of this task lies in the comparison of usability between manual highlighting and chat-assisted highlighting, we have not asked the participants to write down the names of the highlighted classes.
After asking the chat assistant to complete the same task, all participants rate the correctness of the tool actions as very good and state that no mistakes were made.
By inspecting at the log files of chats, we can confirm that the chat assistant found all six model classes that were present in the PetClinic.
When comparing the ease of use between chat assistant and manually highlighting classes, all participants preferred the chat assistant.

In Task 5, the participants should compare the manual creation of color themes with the creation of color themes with the chat assistant.
For two participants, the color theme creation was not satisfactory with one participant stating that he/she ``asked for a rainbow and got a blue background and shades of orange'', while the other participant stated that it ``seems like most colors weren't adjusted at all''.
This is an issue with the reliability of the chat assistant. 
However, the ease of use for changing color themes was rated highly, averaging at 4.82/5 points.

In Task 6, the participants were asked to use the chat assistant for a visual software restructuring.
Only the expectations of four people were fully satisfied or satisfied (see \autoref{fig:evaluation}), resulting in a mean value of 3.0/5 points for matching the participant's expectation.
This is also reflected in the written comments:
\begin{itemize}
    \item ``asked for to improve the communication arrows but the result was a bit off''
    \item ``It has to be motivated to do so. It got really confused and did not remove the old classes''
    \item ``I asked the chatbot for another landscape ordering (all applications in one line), but nothing has changed. I checked the settings and the Application Layout Algorithm said `Horizontal' but after checking the dropdown menu, that item doesn't exist.''
    \item ``Asked to implement service described earlier in the conversation when experimenting with the bot after task 2. First the bot failed. When told again with `you can do it.' it succeeded.''
    \item ``The chat bot lied to me! ;) After convincing the chatbot to try its best for lining all applications up, it has tried its best and the landscape actually changed''
\end{itemize}

The quality of the explanations and the overall perceived correctness/quality of the responses were rated with 3.73/5 and 3.18/5 points, respectively.
We also asked the participants if they were asked to confirm the changes before the chat assistant applied them to the software landscape.
For eight out of eleven participants, an additional prompt or confirmation was required.
One exemplary reason for the confirmation is that ``The agent needed additionally information about how to improve the metrics''.

After the tasks were completed, the participants were asked to rate their overall experience with the chat assistant.
Regarding response time, nine out of eleven participants rated it as fast or rather fast (see \autoref{fig:evaluation}), averaging at 3.91/5 points.
The participant that gave the lowest rating for response times stated: ``It is fairly slow given that it doesn't have to query data from anywhere except what is already present in the frontend, but maybe this is to be expected when working with an LLM''.
Another participant who rated the response time as rather fast added: ``Fast except for when generating new components''.

Regarding the overall ease of use, all participants rated it as easy to use with one participant commenting: ``Very intuitive to use, given that it follows the prompts''.
In the last question, the participants should rate their overall experience with the chat assistant, resulting in four people who rated it as very good and seven people who rated it as good.
Numerically, this leads to an average score of 4.36/5 for the overall experience.

As the created Claude API key was only used for our study, we could retrieve data on the token usage and cost from the Claude Console.
The results are presented in \autoref{tab:tokens}.
Although we do not have statistics on token usage by individual participants, the collected chat logs suggest that the number of used tokens is not distributed evenly among participants.

\begin{table}[h]
\centering
\begin{tabular}{|c|c|}
\hline
\textbf{Metric} & \textbf{Value} \\
\hline
Model & Claude Sonnet 4.5 \\
\hline
Total Input Tokens & 11,401,606 \\
\hline
Total Output Tokens & 135,431 \\
\hline
Total Cost & \$36.24 \\
\hline
\end{tabular}
\caption{Token usage and resulting cost according to Claude Console for employed Claude Sonnet 4.5 model.}
\label{tab:tokens}
\end{table}

\subsection{Discussion}
In this subsection, we discuss the results with regard to our three main areas of interest: program comprehension, tool execution, and chat-based software restructuring.

\textbf{Program Comprehension} 
With regard to program comprehension, we could observe that all participants reported having gained additional information by prompting the chat assistant.
As the chat assistant is integrated into the frontend of the visualization tool, the collected data that is the basis of the visualization can be directly used by the chat assistant.
However, without explicitly stating how detailed the answer should be, the responses seem to be very detailed, and for some participants, too detailed.
This could be improved by adding instructions to the chat assistant that it should generate brief responses by default.
In addition, our chat interface should be improved to account for longer responses, as one participant noted about his overall experience that the chat ``window could be larger to see more text at once reducing the scrolling''.
The good feedback on response times and ease of use indicates a high usability of our chat assistant for program comprehension tasks.
A clear drawback is the unreliable correctness of the given responses.
Individual participants noted this fact or reported actual hallucinations.
To counteract this, the chat assistant would need to include verifiable sources for all statements in its response.

Overall, we conclude that chat assistants are a promising tool to improve program comprehension in software visualization tools.
Advantages are the high ease of use, high flexibility, and relatively low response times (depending on the employed LLM).
Disadvantages are the unpredictable character of given responses regarding their content, length, and correctness.

\textbf{Tool Execution}
The tool calls of deterministic and well-defined features, such as highlighting and color theme creation, consistently received high ratings for their perceived correctness and ease of use.
This suggests that triggering changes in a software visualization through natural language with a chat assistant can reduce interaction overhead, particularly for operations that would otherwise include multiple steps in the user interface.
However, this approach does not eliminate all manual efforts: the result of requested tool actions still needs to be verified to ensure they were indeed executed correctly.
In addition, the natural language interface allows for any sort of prompts and not all can be resolved to the user's satisfaction.
For example, one participant who asked the chat assistant for a rainbow color theme would have wished for the background of ExplorViz to be more colorful even though it can only be set to one color.
This phenomenon also arose when one participant asked the chat assistant to align all applications in a row, something that the chat assistant has no direct control over and which led the chat assistant to apply the hallucinated value ``horizontal'' to the layout settings.

Overall, we conclude that chat-assisted tool execution can reduce the perceived effort and improve the usability compared to manual interactions with the user interface of a software visualization tool.
Although the perceived correctness was relatively high in our evaluation, we deem it of high importance to have strong guardrails for the provided frontend tools, as well as do further prompt engineering and be able to detect and react to faults of the chat assistant when executing tool calls.

\textbf{Chat-Based Software Restructuring}
The software restructuring was rated lower and exhibited greater variance among participants compared to the tool execution in previous tasks.
The chat assistant asked most participants for more detailed information or confirmation for the planned changes.
This illustrates the complexity that comes with software restructuring tasks.
It may be that improved prompting with more detailed requirements for the software restructuring would have significantly improved the chat assistant's tool calls.
However, we argue that integration of tools for chat assistants for complex software engineering use cases that go beyond simple updates of the visualization needs special attention in the implementation.

Advantages still lie in the ease of use of a natural language interface that can trigger complex changes to the visualization.
Overall, we conclude that complex visualization updates such as software restructuring workflows require stronger guardrails for the chat assistant and clearer previews and improved control mechanisms for the user.

\textbf{Token Usage (Costs)}
Regarding the token usage, it can be clearly seen that many more input tokens have been used than output tokens (roughly 84 times as many).
This shows that the most potential to reduce cost for our approach lies in reducing the input token usage.
This could be done by more fine-grained access to the available visualization state for the chat assistant and improved instructions on when to retrieve what kind of context.
In addition, previous chat messages are sent as context alongside new prompts.
For this, automatic summarization of chat messages or clearing the chat completely could have a significant impact on the token usage.
It should be noted, though, that the cost of input tokens is usually significantly lower than the cost of output tokens.\footnote{\url{https://platform.claude.com/docs/en/about-claude/pricing}}

\subsection{Threats to Validity}
The participants of our evaluation were mostly computer science students and thus are not representative of the group of professional software developers.
The relatively small sample of eleven people limits the significance of our results.

Some of the participants knew the study's conductor personally, and they were offered the chance to be compensated with a €15 gift card.
In addition, a number of participants have already worked on ExplorViz or have taken part in earlier evaluations of ExplorViz.
Therefore, some participants may already be familiar with the Spring PetClinic, which could result in contamination of the training data.
Due to previous exposure to ExplorViz and other studies, participants may have been influenced by demand characteristics, leading them to believe that the chat assistant is superior to manual use of ExplorViz.

The employed software landscape and tasks were designed to foster interaction with the visualization and our implemented chat assistant, without the need for prior knowledge about ExplorViz.
Thus, we did not consider complex and arguably more realistic tasks that include more in-depth software engineering activities on the employed software landscape.

Since the manual version of the task was always completed before the chat assistant was used, ordering effects may have influenced the responses due to fatigue, practice effects, or the carryover of knowledge.

With Claude Sonnet 4.5 we employed only one of many available LLMs.
In the meantime, updated and more expressive models are available, too.
The extent to which the gathered results can be transferred to other LLMs is unclear.

%% file: chapters/conclusions.tex
\section{Conclusions and Future Work}\label{sec:conclusion}
This work explored integrating a chat assistant into 3D software visualization tools such as ExplorViz to improve program comprehension and reduce interaction overhead.
The implemented chat assistant can request contextual data from the visualization and use tool calls to trigger changes to it.
Via this, the chat assistant can request data about the visualized software system such as package hierarchies and trigger actions such as highlighting visual entities or changing the structure of the visualized software system.
A modular backend service connects the assistant to various model providers.

We conducted an evaluation involving eleven participants who assessed the assistant's usability in tasks related to program comprehension, action execution, and the special case of software restructuring.
Participants generally found the generated summaries and explanations accurate and the tool calls, triggered by natural language, easy to use.
These findings suggest that chat assistants could effectively complement interactive software visualization tools, particularly when operating within well-defined operations and provided with relevant context about the visualization.

However, the study highlighted limitations when the assistant was employed for open-ended tasks such as executing software restructuring actions in the visualized software system.
Participants' ratings for expectation alignment in editing were notably lower, indicating that such software restructuring workflows require stronger guardrails for the chat assistant and clearer previews and improved control mechanisms for the user.
Overall, the assistant was considered usable and promising, but its benefits hinge on its grounding in application state and the predictability of its actions.

We would like to stress that our evaluation should only be seen as an indication that the field of combining software visualization and artificial intelligence is promising.
More research should be conducted to explore the integration of chat assistants in different kinds of software visualization tools.
Also, quantitative results that compare effectiveness of chat assistants to manual user interactions with industry professionals are highly needed.
As our evaluation could only explore the short-term use of chat assistants, the long-term use of such tools where the chat assistant could adapt to the user's observed inputs and preferences would be very interesting.

Regarding the implementation of the chat assistant, we see further potential in optimizing instructions, prompts, provided context, and tool calls to improve the chat assistant's quality of responses and reduce token usage.
The intelligent and task-dependent combination of different LLMs could play a crucial role here in the future.